\documentclass[a4paper,preprint,superscriptaddress]{revtex4}
\usepackage{graphicx,color}
\usepackage{url}
\usepackage{amssymb}
\begin{document}
\title{Confined colloidal crystals in and out of equilibrium}
\author{A. Reinm{\"u}ller}
\affiliation{
Institut f\"ur  Physik,
Johannes-Gutenberg-Universit\"at Mainz,
55128 Mainz, Germany}
\author{E. C. O{\u{g}}uz}
\affiliation{
Institut f\"ur Theoretische Physik II: Weiche Materie,
Heinrich-Heine-Universit\"at D\"usseldorf,
40225 D\"usseldorf, Germany}
\author{R. Messina}
\affiliation{
Institut de Chimie, Physique et Mat\'eriaux (ICPM),
Universit\'e de Lorraine,
1 Bd. Arago, 57078 Metz - Cedex 3, France}
\author{H. L{\"o}wen}
\affiliation{
Institut f\"ur Theoretische Physik II: Weiche Materie,
Heinrich-Heine-Universit\"at D\"usseldorf,
40225 D\"usseldorf, Germany}
\author{H. J. Sch{\"o}pe}
\affiliation{
Institut f\"ur  Physik,
Johannes-Gutenberg-Universit\"at Mainz,
55128 Mainz, Germany}
\author{T. Palberg}
\affiliation{
Institut f\"ur  Physik,
Johannes-Gutenberg-Universit\"at Mainz,
55128 Mainz, Germany}

\begin{abstract} 
Recent studies on confined crystals of charged colloidal particles are reviewed,
both in equilibrium and out of equilibrium. We focus in particular on  
direct comparisons of experiments (light scattering and microscopy) 
with lattice sum calculations and computer simulations.
In equilibrium we address buckling and crystalline multilayering of charged systems in hard and 
soft slit confinement. We discuss also recent crystalline structures obtained for charged mixtures. 
Moreover we put forward possibilities to apply external perturbations, in order
to drive the system out of equilibrium. These include  electrolyte gradients 
as well as the  application of shear  and  electric fields. 
\end{abstract} 
\maketitle
\section{Introduction}
\label{sec:intro}
Under strong confinement,  as realized by system boundaries or laser-optical external fields, the freezing transition of a colloidal suspension is not only significantly shifted relative to its bulk freezing point \cite{Loewen_1994} but the crystalline structure itself can differ much from its stable bulk phase \cite{Lowen2012_Advances,Schoepe2006_Langmuir,Fontecha_2007,Fontecha2008_PRE,apolinario2006inhomogeneous,brunner2002phase,reichhardt2002novel,fortini2006,Qi2013_arXiv,Rice2009_ChemPhysLett,Erdal_JPCM}. An example of very strong confinement is given in a slit--geometry between two parallel plates which results in a few crystalline layers if the plate distance is small \cite{murray1990comparison,Erdal_2009_EPL}. The extreme limit is a quasi-two-dimensional colloidal monolayer. Upon widening the confinement, the latter can buckle into a bilayer \cite{pieranski1983thin,Peeters_PRL_1999,Messina_2003,Fontecha2005_JPCM,Schoepe2006_Langmuir,fortini2006} and subsequent multilayered structures do emerge. Recent research has focused on the structural details of the resulting multilayered crystal. Colloids are excellent model systems to observe these structures on the particle scale using optical microscopy as they can be confined between glass plates. Moreover, various kinds of external fields can be used to bring the system out of equilibrium \cite{Lowen2001,Villanova-Vidal2009_PhilosMag,Schoepe2003_JCPM,Marques-Huesco2008_incoll} resulting in novel phenomena not seen in equilibrium or to study the dynamics of the crystallization process in detail.

The nature of the confinement can most easily be described by an external one-body potential acting on the colloidal particles, for recent examples see \cite{chakrabarti1995reentrant,bechinger2001phase,strepp2002phase,Resnick2003_JColloidInterfaceSci,Brunner2004_PRL,apolinario2006inhomogeneous,brunner2002phase,reichhardt2002novel,mikhael2008archimedean,schmiedeberg2008colloidal,Schmiedeberg2007_EPJE}. One can therefore distinguish between hard and soft confinement: Hard confinement implies a hard external potential, i.e. a potential which is either zero or infinity describing just a steric volume exclusion of the particles. Such an external interaction is athermal, i.e. if scaled with the thermal energy, it is the same at different system temperatures which can reduce the thermodynamic parameters of the total system. Soft confinement, on the other hand, is described by a smooth external potential. An important example are charged glass plates which act as a soft confinement to charged colloids as described by an exponentially screened Coulomb interaction \cite{hansen2000,Lowen2008_JCPM,Oguz_2012}.

The goal of  this minireview is twofold. First, we summarize some recent progress in the experimental determination and theoretical prediction of colloidal multilayers in various kinds of hard and soft confinement under equilibrium conditions. As a result, the stable equilibrium structures reveal a fascinating complexity being very sensitive to both particle interactions and kind of confinement. Most of the structures can be understood by a simple potential energy minimization, i.e. on purely energetic grounds. Second, we focus on nonequilibrium situations which originate from the application of an external field, like gravity, shear flow and an electrolyte gradient. We show that, under appropriate solvent flow conditions, a seed can generate the controlled growth of a colloidal monolayer. Moreover, we discuss how gravity and electric fields can change crystallization. Last but not least, we describe a complex self-organization process of colloidal clustering into a swimming object in an electrolyte gradient. The physical reasons underlying the details of the  swimming process still need to be understood and clarified in future studies.

The minireview is organized as follows: in chapter 2, we summarize all aspects of confined colloidal crystal in equilibrium. Nonequilibrium effects as induced by different external fields are subsequently discussed in chapter 3 and we conclude in chapter 4.

\section{Phase behaviour of confined colloidal crystals in equilibrium}
\label{sec:phase_behaviour_eq}
\subsection{Charged colloids in soft confinement}
\label{subsec:ChargedSoftConfinement}
Regarding crystallization in slit geometry, hard confinement has been studied for both hard particles and soft particles i.e. particles interacting with a soft pair potential like a Yukawa interaction. More recently, the case of hard particles in soft confinement has also been addressed \cite{Curk2012_PRE}. Complex cascades of multilayered crystals were predicted and confirmed in colloidal experiments the details of which depend on the interparticle and particle--wall interactions. Detailed comparisons between experiment and theory were performed for hard and soft particles in hard confinement but for soft particles in soft confinement such a direct comparison was done only recently \cite{Oguz_2012}, which we describe in more detail in the following.

A flexible realization of soft sphere systems is provided by charged colloids suspended in water. The screened Coulomb interaction can conveniently be varied between the theoretical limits of hard spheres (approached at high salt concentrations) and the One Component Plasma (OCP, approached at low salt concentrations).
To investigate charged colloidal spheres under deionized conditions in confinement between parallel flat walls, Reinm\"uller et al. realized a new experimental set--up to be used in combination with a commercial scientific microscope \cite{Reinmueller2013_RevSciInst}. An experimental set--up for flexible and accurate investigations of such systems under confinement requires a number of minimum properties: It must allow for fast and precise adjustment of the confining geometry; it must facilitate efficient preparation of colloidal suspensions under low salt concentrations at absence of unwanted chemical gradients; it must provide an efficient homogenization mechanism as well as long term stable experimental conditions. The measurement cell of the setup consists of optically flat quartz substrates attached to piezo actuators providing a fast and flexibly adjustable confining geometry. The local wall separations and thus the confining geometry can be quantitatively controlled in situ by interferometric measurements. Proper choice of materials guarantees sufficient chemical inertia against contamination of any kind. Sample preparation and homogenization is performed efficiently using a syringe pump in combination with a conventional closed tubing system including a mixed bed ion exchanger column. Combining optical microscopy, conductivity measurements as well as interferometric measurements, the setup facilitates quantitative control of particle interaction parameters and confining geometry as well as the determination of the structural and dynamic properties of the confined colloids.

\begin{figure}
\centering
\resizebox{1.0\columnwidth}{!}{\includegraphics{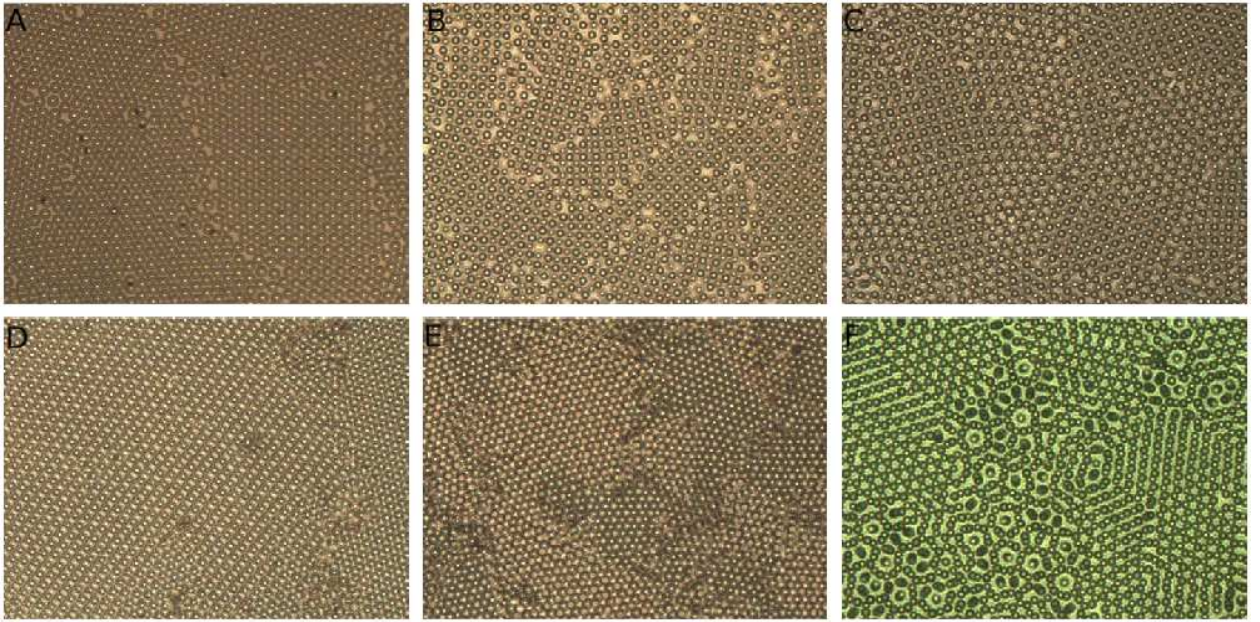}}
\caption{Examples of crystaline structures observed in experiments on charged spheres in soft parallel plate confinement: $1\triangle$ (A), $2\square$ (B), coexistence of $2\square$ and $2\triangle$ (C), $2\textrm{hcp}\perp$ (central large domain) in coexistence with $2\triangle$ (grain at the right margin)(D), $3\triangle$ with different appearances of fcc(111) and hcp(001) faces (E) and Moir\'e rosettes in coexistence with $2\textrm{hcp}\perp$ and $2\triangle$ (F). (Field of view: $280 \times 210 \mu\textrm{m}^2$.)
From Ref.\ \cite{Oguz_2012}.}
\label{fig:1}   
\end{figure}
\begin{figure}
\centering
\resizebox{0.6\columnwidth}{!}{\includegraphics{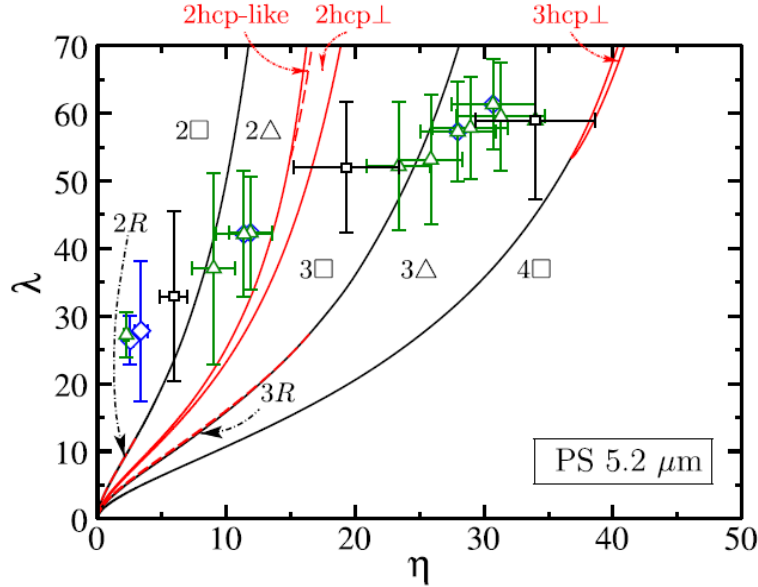}}
\caption{Comparison of the theoretical prediction (dashed and full
lines) for the multilayer stability phase diagram at zero temperature to the experimental data (symbols) taken from the phase diagram for PS $5.2 \mu \textrm{m}$ particles in \cite{Oguz_2012}. We use $\kappa a = 6$ and a particle/wall surface charge ratio of $\nu = 0.4$, where $a$ is the particle radius. From Ref.\ \cite{Oguz_2012}.}
\label{fig:2}   
\end{figure}

Two mono-disperse aqueous suspensions of negatively charged polystyrene spheres with diameters of 5.2$\mu$m and 2.6$\mu$m (batch nos PS/Q-F-B1036 and PS-F-B233 by MicroParticles Berlin GmbH, Germany) were used to explore the equilibrium phase diagram of charged colloidal spheres in aqueous suspensions under spatial confinement in parallel plate geometry at low background salt concentrations using the home-made setup described above. In order to suppress gravity, the solvent mass density was matched to that of polystyrene by adding 20 vol\% glycerol. During the  experiments, the chemical properties of both particle and substrate surfaces were kept constant. The equilibrium phase diagram was
determined in terms of the emerging crystal structures depending on the dimensionless parameters the reduced area number density $\eta=n_A d^2$ and the reduced inverse screening length $\lambda=\kappa d$, where $n_A$ denotes the area particle number density, $\kappa$ is the inverse screening length, and $d$ is the width of the confining slit. These parameters could be determined by accurate measurements of the area number density, the cell height, the salt concentration and the particle charge. Varying the reduced area number density $\eta$ and the reduced inverse screening length $\lambda$ in the experiment, the sequence of crystalline structures shown in Figures~\ref{fig:1} and \ref{fig:2} was observed. Basically these are hexagonal closed packed crystals (fcc or hcp) oriented in different crystallographic directions with respect to the cell walls. There are no prism structures, transient structures (e.g. hcp like) or exotic structures, like the Belgian waffle iron structure \cite{Erdal_2009_EPL}, observable.

The experimentally observed equilibrium phase behavior of highly charged colloidal crystals in parallel soft confinement was found to be in very good agreement with the theoretical ground state phase diagrams of expponentially screened point-like Coulomb (Yukawa) particles calculated by lattice sum minimizations. The particles were exposed to a soft confining hyperbolic cosine potential as described by the linear screening theory \cite{Andelman}.  The phase diagrams display the same simple multilayer phase sequence with increasing system density. The agreement is documented in Figure~\ref{fig:2} which contains a comparison between experiment and theory in the two--dimensional plane spanned by $\eta$ and $\lambda$ for the $5.2 \mu \textrm{m}$ size colloids.

\subsection{Hard slit confinement}
\label{subsec:hard_slit_confinement}
One of the simplest situation of confined colloids are sterically-stabilized particles ("hard spheres") 
confined between parallel hard walls. In this slit geometry the packing fraction and the reduced plate distance 
$L/\sigma$ (with $\sigma$ denoting the hard sphere diameter) are the only thermodynamic parameters \cite{Schmidt1,Schmidt2}. 
The two-dimensional limit of extreme confinement, $L \to \sigma$, gives a system of hard disks. In this case, the close-packed situation is a hexagonal (or triangular) two-dimensional crystal. For finite packing fractions, the nature of the freezing transition has attracted large attention in the last years \cite{Bernard2011,Peng10,Engel2013_PRE}. 
For slightly larger plate distances, the spheres deviate a bit from their ideal mid-plane positions. In this case,
the perturbative correction to the exact partition function can be calculated analytically \cite{Franosch2012}.
The cascade of close-packing sequences as a function of $L/\sigma$ is nontrivial once the distance 
exceeds that of two intersecting hexagonal layers, see \cite{ivlev2012complex} for a review. 
The phase diagram as determined from free-energy calculations was first investigated in \cite{fortini2006}, which reveals the thermodynamic stability of close-packed structures such as prisms. 
More recently the best packed configuration was found by systematic numerical minimization routines
\cite{Oguz_2012_PRL}. As a result,  periodic
adaptive prismatic structures which are composed of alternating prisms of spheres are best packed. The internal structure
of these prisms adapts to the slit height which results in close packings for a range of plate separations, just
above the distance where three intersecting square layers fit exactly between the plates. The adaptive
prism phases were also observed in real-space experiments on confined sterically stabilized colloids and in
Monte Carlo simulations at finite pressure \cite{Oguz_2012_PRL}.
\subsection{Cylindrical confinement}
\label{subsec:cylindrical_confinement}
If particles are confined in cylindrical tubes, there is no sharp 
freezing transition since phase transitions in general are rounded in one dimension by fluctuations \cite{Wilms}.
However, the ground state at zero temperature can be a  periodic crystal.
It has been shown, that hard spheres confined to a
hollow hard cylinder show complex ordering structures with helical order \cite{Pickett2000}.
The ground state structures were generalized recently to Yukawa systems in a hard cylindrical pore
\cite{Oguz_2011}. As a function of
screening strength and particle density, the equilibrium phase diagram was found to exhibit a cascade of
stable crystals with both helical and non-helical structures.
\subsection{Monolayers of oppositely charged colloids}
\label{subsec:monolayers}
\begin{figure}
\centering
\resizebox{0.6\columnwidth}{!}{\includegraphics{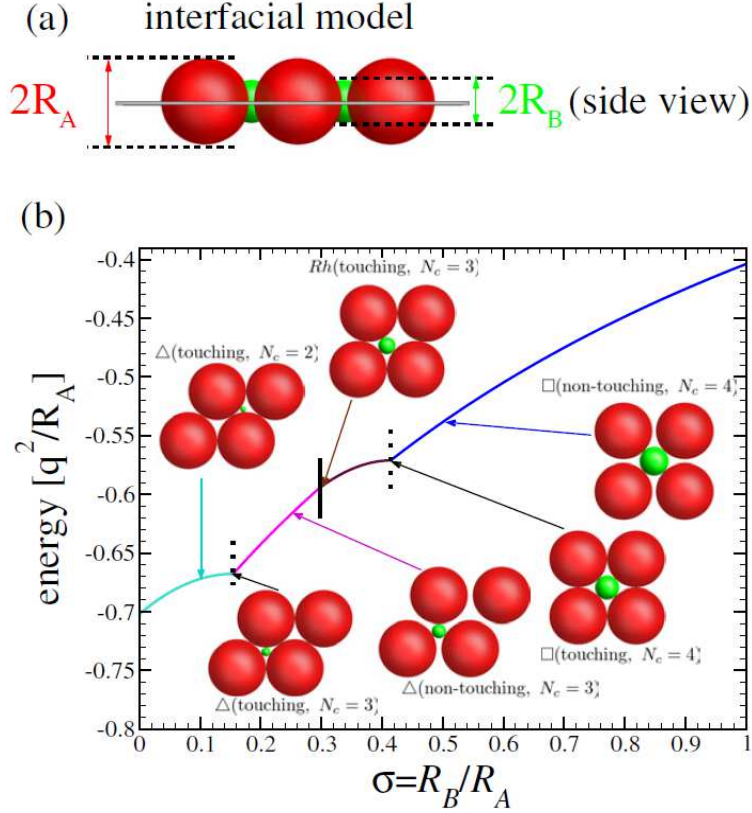}}
\caption{Stable structures of oppositely charged
spheres versus their size asymmetry $R_{B}/R_{A}$ in the interface
model, where all sphere centers fall on the same plane: a) side
view, b) (scaled) energy per ion ($q$ is the particle charge). 
The discontinuous transition is
indicated by a solid bar. Continuous transitions are denoted by
a broken bar. Unit cells of the corresponding stable phases are
shown, where the big (small) spheres have a radius $R_{A}$ ($R_{B}$).
From Ref.\ \cite{Assoud2010_EPL}.}
\label{fig:3}   
\end{figure}

Oppositely charged colloids form stable crystals even at zero pressure due to the mutual attractive Coulomb forces, for a recent review see \cite{ivlev2012complex}.
A model of oppositely charged hard spheres with different diameters is a standard description for the basic interactions.
In three dimensions the ground state at zero pressure is textbook knowledge \cite{evans1966crystal41to43}. There are three
basic equimolar structures whose stability depends on
the ratio of the ion radii involving the cesium--chloride, sodium--chloride, and zincblende structures.
These (and some more) structures are not only realized for molecular salts but also for mesoscopic colloids \cite{leunissen2005,hynninen2006,hynninen2006cuau}. 
In Refs. \cite{Assoud2010_EPL}, the same problem was solved for two-dimensional monolayers of oppositely charged hard spheres. 
Among the stable structures are square, triangular and rhombic crystals as
well as “empty” crystals made up of dipoles and chains, which have a vanishing number density.
The square structure was already observed for charged granulates \cite{Kaufman2009}. 
As an example, Figure~\ref{fig:3} shows the cascade of zero-pressure ground states as a function of the diameter ratio
for an interface set--up (interfacial model) where the centers of mass of the two species are all in a common plane. The associated ground 
state energy per particle is plotted versus the diameter ratio. 

\section{Phase behaviour of confined colloidal crystals out of equilibrium}
\label{sec:phase_behaviour_noeq}
\subsection{Influence of gravity}
\label{subsec:Influence_of_gravity}
The influence of gravity on the crystal formation in a two layer system of charged colloidal particles in confinement was explored experimentally using commercial polystyrene spheres of 5.2 $\mu$m in diameter in fully deionized water \cite{Moire}. Here, the colloidal bilayer crystal assembly occurs under influence of both gravity and electrostatics.
During a first stage of settling, bilayers form via a line pattern, which quickly gives way to a polycrystalline bilayer structure of AB--stacking of two triangular monolayers. Due to gravity, this bilayer structure is a unidirectional distorted crystal: The lattice constant in vertical direction is noticeable smaller than in lateral direction.
In addition to the triangular lattice, two extraordinary transient structures were identified. Each is made from two perfect triangular layers which are rotated by certain angles with respect to each other. The intergrowth of the conventionally stacked two layer crystals results in $1 \times 1 R \alpha$ super structures. Two specific rotation angles $\alpha = 27.8^\circ$ and $\alpha = 38.2^\circ$ dominate in the observed two layer system. These superstructures show fascinating triangular Moir\'e patterns of rosettes. The morphological function of these structures can be characterized as an ``extended grain boundary'' known from atomic systems. A similar coherent intergrowth of crystal domains in the bulk are observed e.g. in rare earth boride carbides.  Due to their complex structure, the observed crystalline arrangements display extraordinary scattering patterns showing a twelve-fold symmetry. Measurements and model calculations demonstrate the high potential of these extraordinary structures for photonic applications.
\subsection{Crystal formation in parallel confinement under the influence of shearing fields}
\label{subsec:crystal_formation_shearing}
Regarding crystallization in confined geometry, heterogeneous nucleation at the cell wall and the influence of shearing fields are of substantial importance. Colloidal crystals are easily shear molten by gentle mechanical agitation. After cessation of shear, they readily crystallize via nucleation and growth.  
Applying randomly directed shear before crystallization bcc crystals nucleating at the wall are oriented with their (110)-bcc plane parallel to the wall, but do not show any preferred orientation of the (111)--direction \cite{Engelbrecht2011_SoftMatter}. The crystal growth in vertical direction was observed to be reaction controlled with a constant growth velocity and as function of meta-stability the growth velocities follow a Wilson-Frenkel-law. The resulting polycrystalline material consist of columnar crystals randomly orientated in the crystallographic direction parallel to the cell wall.

Analyzing the heterogeneous nucleation rate densities on the wall as function of metastability, a transition from a nearly time independent nucleation rate density at the phase boundary to a peaked transient nucleation at higher undercooling $\Delta \mu$ was identified. The heterogeneous nucleation barrier height stays first constant and drops down to zero as the metastability increases: In charged sphere model systems, a sharp wetting transition separates a regime at moderate metastability where hemispherical cap nuclei are formed from a strong meta-stability regime, where the crystal phase completely wets the wall.

In suspensions molten by unidirectional shear flow this, transition could not be observed. Unidirectional shear leads to shear induced oriented precursor crystals in form of hexagonal layers at the cell wall ((111)--fcc parallel to the wall) and further quasi-epitaxial growth of bcc crystals on this precursor crystal \cite{Wette2009_JCPM,Palberg2012_JCP}. The precursor acts as structured substrate reducing the nucleation barrier height to a minimum. The details of this first step are still not well understood. Due to the unidirectional shear, the bcc crystals are uniformly oriented ((110) plane parallel to the wall and (111)--direction in direction of shear flow) leading to a twinned bcc morphology. Like in the first case, vertical growth is reaction controlled with a constant growth velocity. 
In lateral dimension, the average domain size was observed to increase significantly during vertical growth following the typical power law for coarsening with an exponent of 0.5. Lateral coarsening ceases when the wall crystals from the opposing walls intersect. 

These investigations and others \cite{Messina_2006} show impressively that the heterogeneous nucleation process on the wall, the crystallization process afterwards and the crystalline microstructure is highly modified by shear in confined geometry. 
Finally, rigid colloidal spheres rotate in shear flow. Flexible neutral polymer chains, on 
the other hand, tumble \cite{He_2010}. If combined with other periodic field
there are interesting dynamical modes of polymer propagation  \cite{He_2011}.
The behaviour is expected to be even more complex for charged polymers \cite{Allahyarov_2011}.

\subsection{Crystal and microswimmer formation in parallel confinement under the influence of electrostatic and flow fields}
\label{subsec:crystal_microswimmer}
\subsubsection{Crystal nucleation}
\label{subsubsec:Crystal_nucleation}
\begin{figure}
\centering
\resizebox{1.0\columnwidth}{!}{\includegraphics{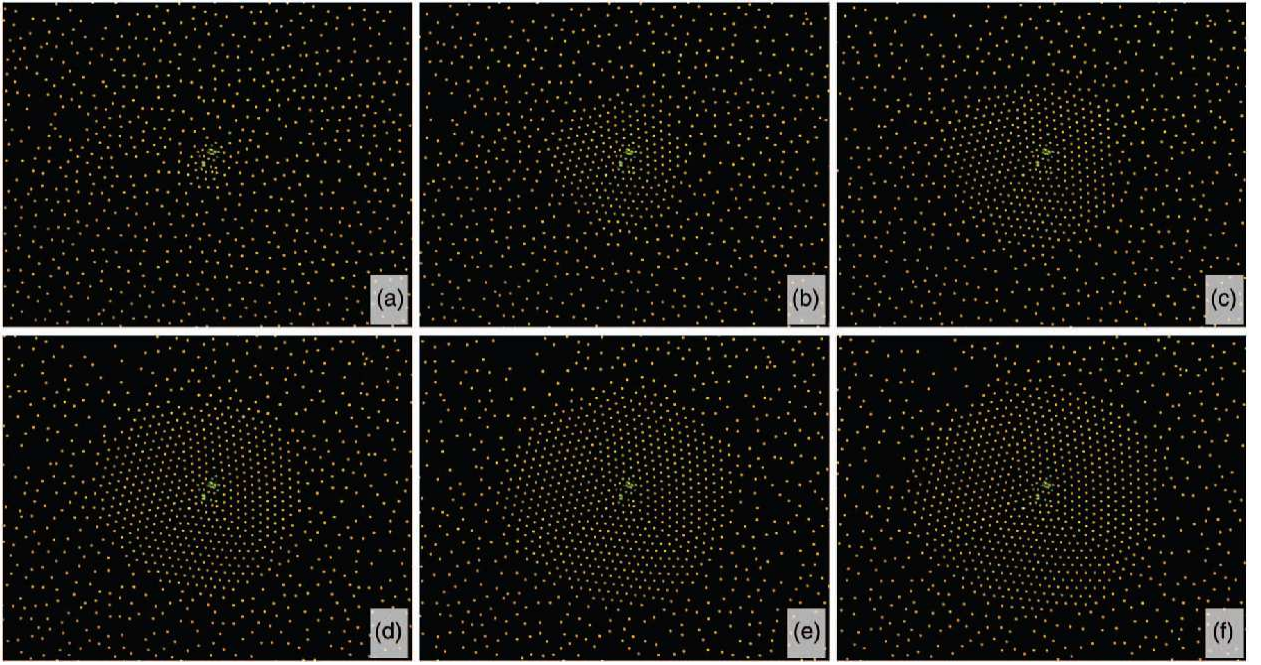}}
\caption{Growth of a three--domain monolayer crystal on a seed at $t = 0 \textrm{s}$ (a), $t = 100 \textrm{s}$ (b), $t = 200 \textrm{s}$ (c), $t = 300 \textrm{s}$ (d), $t = 400 \textrm{s}$ (e), and $t = 500 \textrm{s}$ (f); field
of view: $360 \times 290 \mu \textrm{m}^2$.
From Ref.\ \cite{Reinmueller2012}.}
\label{fig:4}   
\end{figure}

The crystallization process in equilibrium two--dimensional colloidal fluids in confinement induced by local electrolyte gradients was studied experimentally using negatively charged polystyrene spheres (diameter $2a = 5.2 \mu \textrm{m}$) dispersed in water \cite{Reinmueller2012}. The suspension was prepared under deionized conditions in confined parallel plate geometry obtaining a fluid monolayer system in thermodynamic equilibrium. After introducing small fragments of ion exchange resin into the confinement cell, a strong interaction between the colloidal spheres and the fragments could be observed. In particular for the case of cation exchange resins (CIEX) a fluid flow leads to particle accumulation at the CIEX-fragment followed by colloidal crystallization. The electrolyte gradient originated by the CIEX-fragment caused an electro-osmotic solvent flow along the anionically charged substrate surface towards the fragment leading to a colloidal particle transport. Once the local particle number density exceeded the salt concentration dependent freezing density, heterogeneous crystallization on the surface of the CIEX-fragment occurred. Although the underlying mechanism generating the flow field is not fully understood, the phenomenology of the crystal formation was observed to be well reproducible. Starting from the thermodynamically stable colloidal fluid multi-domain triangular monolayer crystals were formed with the fragment in the centre, as for experimental snapshots see Figure~\ref{fig:4}. The domain number is highly correlated with the shape of the fragment acting as seed particle. Analyzing the crystal growth, an exponentially decreasing crystal growth velocity was observed and the final crystallite size is determined by the strength and range of the electrolyte gradients. The same situation was simulated using Brownian dynamics computer simulations in two dimensions and good agreement with the experiment was obtained.

\begin{figure}
\centering
\resizebox{0.5\columnwidth}{!}{\includegraphics{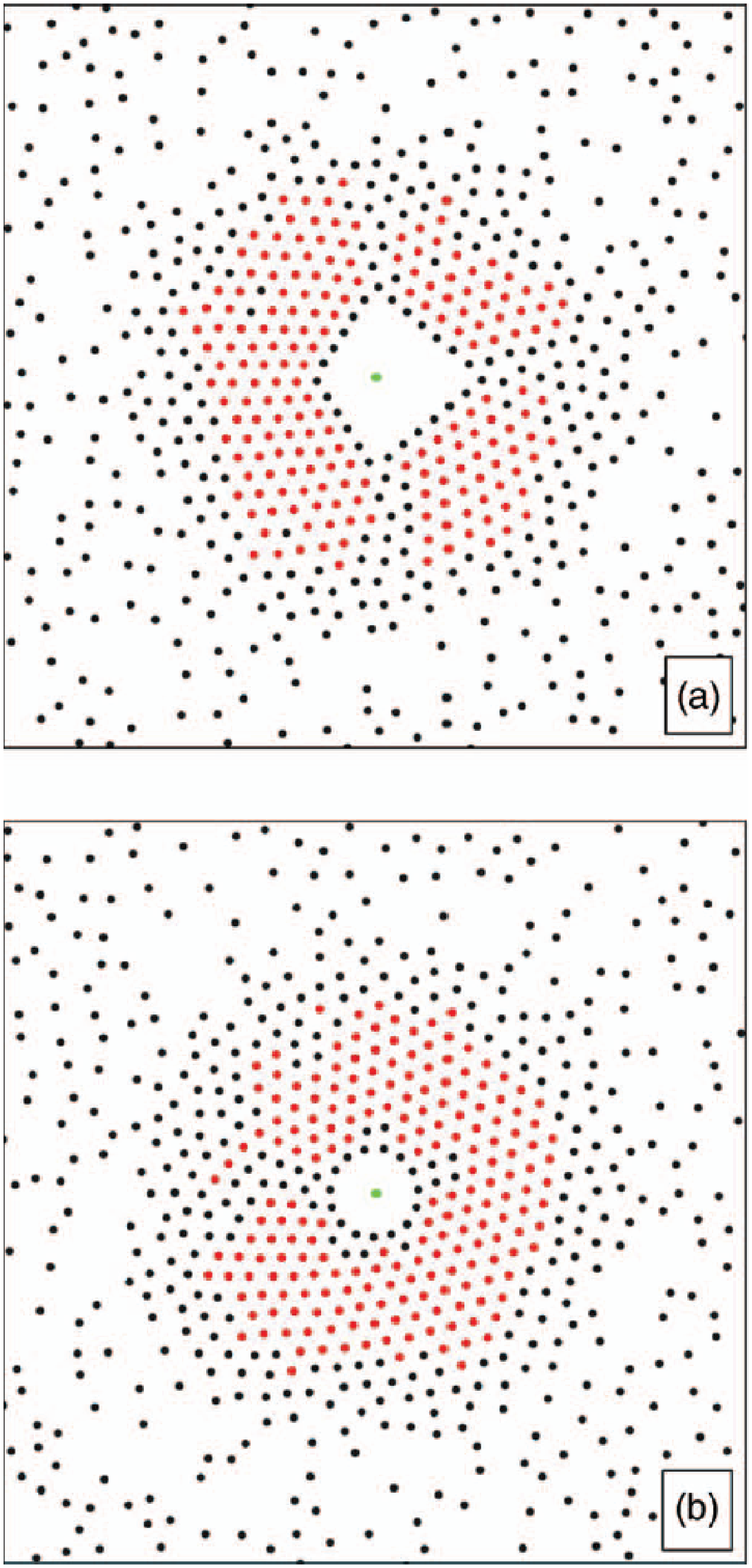}}
\caption{Simulation snapshots of a three--domain (a) and a mono--domain crystal
(b) each at $t = 500\tau$ with $\tau$ being a typical Brownian time scale. The green dots in both illustrations indicate the origin
of the system. The particle flow into the polygonal (a) and spherical (b) area
is prohibited. The red particles are crystalline, while the black
ones are not. Field of view: $250 \times 250/\kappa^2$ where $\kappa$ is the inverse screening length.
From Ref.\ \cite{Reinmueller2012}.}
\label{fig:5}   
\end{figure}

Simulation and experiment indicate that there is no nucleation barrier in this special crystallization mechanism and that the number of domains depends neither on thermodynamics nor on nucleation kinetics. The seed geometry determines the microstructure of the formed crystal, see simulation snapshots in Figure~\ref{fig:5} for two different seed geometries.
In principle these investigations offer the possibility of designing and manufacturing extended thin multidomain colloidal crystal microstructures controlling the positions, shape and size of individual crystals.

\subsubsection{Complex microswimmer formation}
\label{subsubsec:Complex_microswimmer_formation}
Repeating the experiment with stronger electro--osmotic solvent flow and larger plate to plate distance ( $\sim 50 \mu \textrm{m}$)  the formation of self-organized complex microswimmers could be observed \cite{Reinmueller2013_Langmuir}. The electrolyte gradient inducing a diffusio-osmotic solvent flow towards the CIEX fragment now originates a convection roll at the CIEX fragment. Particles transported to the fragment were lifted up at the CIEX and fall down afterward very close to the fragment. Due to the convection the particles did not crystallize, but still a strong accumulation of particles at the fragment was observed. In some cases the particle concentration within the convection cell was found to be highly asymmetric (rotational symmetry, symmetry axis in the centre of the CIEX) which might be generated by either a shape anisotropy of the CIEX fragment or by a highly asymmetric particle concentration of the colloidal fluid surrounding the CIEX fragment.

If the particle concentration within convection cell was highly asymmetric the CIEX fragment and the accumulated particles started to move in the direction of smallest colloid concentration. A possible propulsion mechanism related to the nonuniform colloid concentration (including colloid counterions) might be direct osmotic propulsion. The details of the driving mechanism of this cooperative motion are not yet fully understood. Nevertheless the propulsion was found to be very robust and occurred with varying effectiveness over a large range of experimental parameters. The self propulsion mechanism was self-stabilizing over many minutes and the complex micro swimmers moved with typical velocities of $1 \mu \textrm{m} / \textrm{s} - 3\mu \textrm{m} / \textrm{s}$. At sufficient speed, further loading from the side became negligible and oncoming particles fall down behind the CIEX fragment after they were lifted up by convective flow. During this stage of constant trail thickness, the complexes cover path lengths of many hundreds of micrometers.

\subsection{Oppositely charged colloidal suspensions in electric fields: laning}
\label{subsec:laning}
When oppositely charged colloids are exposed to an external electric field,
laning of like-charge particles occurs, for a review see \cite{ivlev2012complex,Sutterlin09,Dzubiella02,lowen2010instabilities}. 
In three dimensions, by experiment and computer simulation, 
laning was shown to be continuous for small electric fields and a quantitative 
comparison between experiment and simulation could be achieved \cite{vissers2011}.
It is important to note that hydrodynamic interactions between driven particle are of minor importance
when charged particle are moving in an electric field \cite{Ajdari_Long_EPJE_2001,Rex_EPJE_2008}. Therefore
they were not found to be crucial for the actual modelling.

\section{Conclusions}
\label{sec:conclusions}
In conclusion, we have reviewed briefly recent developments in preparing,
characterizing and predicting confined colloidal crystals. In equilibrium,
a huge
variety of equilibrium structures becomes stable depending sensitively on
the particle interaction and on the nature of the confinement. These
structures
may serve as building blocks for filters and sieves as well as for
photonic crystals. A coagulated two-dimensional crystal has controlled void space between 
the particles, which leads to tunable porosity. 
We have also shown that nonequilibrium effects generated by shear flow,
electrolyte gradients and
electric fields may generate further structures which are widely different
from those stable in equilibrium.
These external fields can also be used to steer the crystallization
process itself
and control the defect structure and distribution
of the emerging crystal. One of the most striking consequence was the
self-organized formation
of a microswimming system which opens the way to possible applications as
controlled cargo transport
via active colloidal suspensions.

Future research activities should focus on microscopic theories for
nonequilibrium processes in
order to predict nonequilibrium effects of crystallization.
Dynamical density functional theory is a promising
approach in this respect \cite{seed,Espanol2009,Advances}. On the experimental side,
systems with depletion attractions and binary and ternary mixtures should be explored more. 
In these cases, new segregation and aggregation phenomena are expected which may lead to
further applications. Therefore, though the basic mechanisms are understood by now,  a flourishing 
future of crystallization in confined colloidal suspensions is still lying ahead.

\section*{Acknowledgements}
  This work was supported by the DFG within SFB TR6 (project D1).


\end{document}